\documentclass[conference]{IEEEtran}
\ifCLASSINFOpdf
\else
\fi
\usepackage{algorithm}
\usepackage{algorithm,algpseudocode}
\usepackage{graphicx}
\usepackage{array}
\usepackage{float}
\usepackage{lipsum}
\usepackage{authblk}
\usepackage{authblk}
\usepackage{hyperref}

\usepackage{subcaption}

\begin{document}
%
\title{sOFTDP: Secure and Efficient Topology Discovery Protocol for SDN}
%
%
%

\author[1]{Abdelhadi Azzouni\thanks{abdelhadi.azzouni@lip6.fr}}
\author[2]{Raouf Boutaba\thanks{E.E@university.edu}}
\author[1]{Nguyen Thi Mai Trang\thanks{thi-mai-trang.Nguyen@lip6.fr}}

\author[1]{Guy Pujolle\thanks{guy.pujolle@lip6.fr}}

\affil[1]{LIP6 / UPMC; Paris, France  \{abdelhadi.azzouni,thi-mai-trang.nguyen,guy.pujolle\}@lip6.fr}
\affil[2]{University of Waterloo; Waterloo, ON, Canada  rboutaba@uwaterloo.ca }
\maketitle

\begin{abstract}

Topology discovery is one of the most critical tasks of Software-Defined Network (SDN) controllers. 
Current SDN controllers use the OpenFlow Discovery Protocol (OFDP) as the de-facto protocol for discovering the 
underlying network topology. In a previous work, we have shown the functional, performance and security limitations 
of OFDP. In this paper, we introduce and detail a novel protocol called secure and efficient OpenFlow Discovery Protocol sOTDP.
sOFTDP requires minimal changes to OpenFlow switch design, eliminates major vulnerabilities in the topology 
discovery process and improves its performance. We have implemented sOFTDP as a topology discovery module in Floodlight 
for evaluation. The results show that our implementation is more secure than OFDP and previous security workarounds.
Also, sOFTDP reduces the topology discovery time several orders of magnitude compared to the original OFDP and existing 
OFDP improvements.


\end{abstract}

 {\bf { \it keywords - }}
Software-Defined Networking, OpenFlow, Topology Discovery, security.

%
\IEEEpeerreviewmaketitle

\section{Introduction}


Software-Defined Networking (SDN) introduces the separation between the control plane and the data plane of the network.
SDN moves the control logic to a centralized entity
called the controller. The controller instructs switches/routers in the data-plane as to how packets should be forwarded by 
installing forwarding rules in their forwarding 
tables. To do so, the de-facto standard protocol for communication between the controller and switches is 
OpenFlow. The controller also provides APIs to write network management applications \cite{openflow}.

%

%

\begin{figure}[h]
\centering
   \includegraphics[scale=0.2]{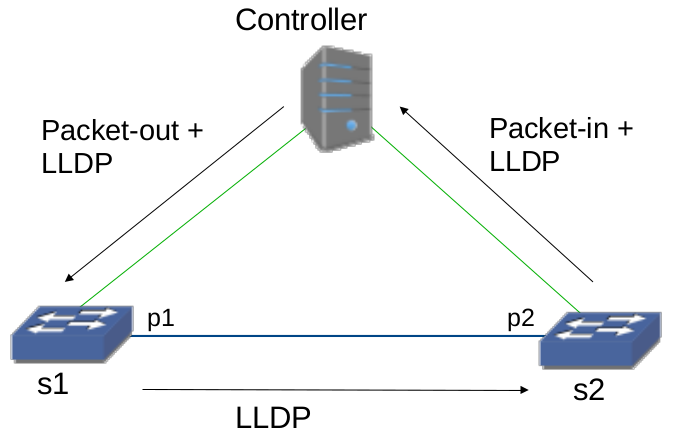}
   \caption{\label{ofdp} Discovering a unidirectional link in OFDP}
   \vspace{-1em}
\end{figure}

One of the controller's duties is to perform an accurate, secure and 
near real time topology discovery to provide management applications with an up-to-date view of the network topology. 
However, all current SDN controllers perform topology discovery using OpenFlow Discovery Protocol (OFDP), which
is far from being secure and efficient \cite{limitations}. 
Figure \ref{ofdp} shows how OFDP works; To discover the unidirectional link s1 $\rightarrow$  s2, 
the controller encapsulates a LLDP packet in a packet-out message and sends it to s1. The packet-out contains instruction for s1 to send the LLDP packet to s2 via port p1.
By receiving the LLDP packet via port p2, s2 encapsulates it in a packet-in message and sends it back to the controller. 
Finally, the controller receives the LLDP packet and concludes that there is a unidirectional link from s1 to s2.
The same process is performed to discover the link in the opposite direction s2 $\rightarrow$  s1 as well as for all other switches in the network.

\begin{figure*}[ht] 
\centering
  \label{ fig7} 
  \begin{subfigure}[b]{0.27\linewidth}
    \centering
    \includegraphics[width=1\linewidth]{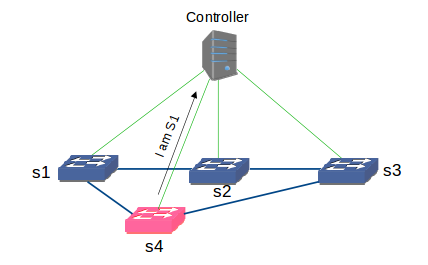}
    \caption{\label{attacks.a}Switch spoofing attack}    
  \end{subfigure}
  \hspace{1em}
  \begin{subfigure}[b]{0.25\linewidth}
    \centering
    \includegraphics[width=1\linewidth]{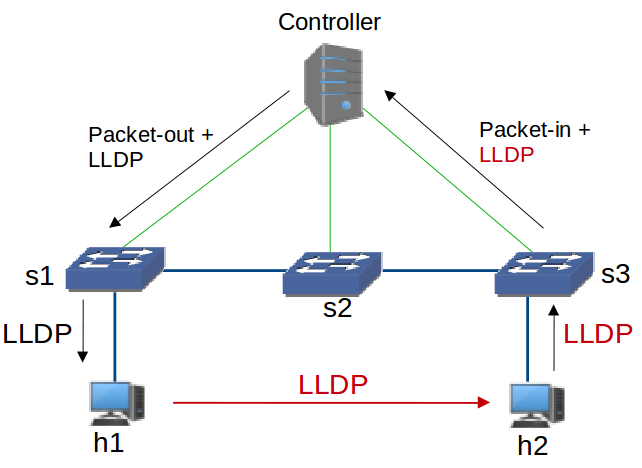}
    \caption{\label{attacks.b}Link fabrication attack}   
  \end{subfigure} 
  \hspace{1em}
  \begin{subfigure}[b]{0.29\linewidth}
    \centering
    \includegraphics[width=1\linewidth]{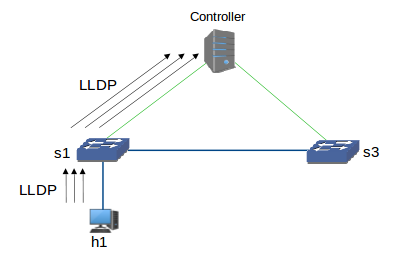}
    \caption{\label{attacks.c}LLDP flooding attack} 
  \end{subfigure}
  \caption{\label{attacks}Attacks on OFDP}
  \vspace{-1em}
\end{figure*}

In highly dynamic networks like large virtualized data centers and multi-tenant cloud networks, 
keeping an up-to-date view of the topology is a critical function; 
Switches join and leave the network dynamically, creating changes in the topology which affects routing decisions that the controller has to make continuously.
To stay up-to-date, the controller needs to repeat the process described in figure \ref{ofdp} periodically. 
The period between two discovery rounds, we will refer to as the discovery period, 
must be chosen carefully based on the dynamicity, size and capacity of the network.
%

An inefficient, vulnerable or buggy topology discovery mechanism can affect routing logic and drastically reduce network performance.
This paper extends our previous work on OFDP limitations \cite{limitations} by detailing the design of a secure and efficient 
alternative protocol that we call $sOFTDP$ (secure and efficient OpenFlow Topology Discovery Protocol). 

The remainder of this paper is organized as follows: In section \ref{whyofdpbad}
we show why the de-facto $OFDP$ shouldn't be implemented in production networks.
We introduce our alternative protocol $sOFTDP$ in section \ref{introducingsOFTDP} and evaluate its performance in section \ref{Evaluation}.
Related work is discussed in section \ref{relatedwork} and section \ref{conclusion} concludes the paper.


\section{Why OFDP shouldn't be implemented in production networks} \label{whyofdpbad} 
In this section, the security and efficiency issues of $OFDP$ are briefly recalled in order for this paper to be self contained.
We refer to \cite{limitations} for a more thorough discussion.

\subsection{OFDP is not secure} \label{attacks}
Current OpenFlow controllers that we have tested (ONOS \cite{onos}, OpenDaylight \cite{odl}, Floodlight \cite{floodl2}, NOX \cite{nox}, POX \cite{nox}, 
Beacon \cite{beacon}, Ryu \cite{ryu} and Cisco Open SDN Controller \cite{opensdn}) implement default $OFDP$ 
using clear and unauthenticated $LLDP$ packets. $OFDP$ is vulnerable to a 
number of attacks including switch spoofing, link fabrication, controller fingerprinting and LLDP flooding.

\textbf{Switch spoofing.} 
LLDP packets in OFDP have two mandatory TLVs: $Chassis Subtype$ and $Port Subtype$ to track packets. 
Most controllers we have tested 
set $chassis Subtype$ value to the MAC address of the local (or internal) port of the switch.
In figure \ref{attacks.a}, the malicious switch s4 
intercepts LLDP packets from s1 containing s1's local port MAC address  
and then use it to connect to the controller as s1.



\textbf{Link Fabrication.} Link Fabrication attacks have two forms: \textit{(i)} Link Fabrication by LLDP relay 
where the adversary has control over two end-hosts h1 and h2 connected to switches s1 and s3 respectively. 
h1 sends the LLDP packets received from s1 to h2 through an out-of-band connection 
(a tunnel over s2 for example), and h2 replicates them to s3 (figure \ref{attacks.b}).
\textit{(ii)} Link Fabrication by fake LLDP injection. If the adversary has control only over h1, but knows the Data Plane 
Identifier (DPID) of s3 then he still can fabricate a unidirectional link between s3 $\rightarrow$ s1 by injecting 
fake LLDP packets into s1 \cite{poisonning}, \cite{insecurity}.
As discussed in \cite{limitations}, authenticating LLDP packets using a static or dynamic key-Hash Message Authentication Code (HMAC) 
does not prevent the link fabrication by relay attack.

\textbf{Controller fingerprinting.} The LLDP packets' content and frequency is different from one controller to another, which
allows fingerprinting attacks on SDN controllers. An adversary (h1 in figure \ref{attacks.b}) matches the LLDP content he/she receives 
from s1 (LLDP packets originating from the controller) against a controller signature database to detect which controller is managing the network. Such information is 
very useful to launch specific and precise attacks on the controller \cite{sdnfingerprint}. 

%

\textbf{LLDP Flooding.} An adversary can exhaust the controller's resources and congest links connecting 
switches to the controller by generating enough fake LLDP packets.
In figure \ref{attacks.c}, h1 generates a large number of LLDP packets and sends them to s1 which has a rule to forward 
every LLDP packet to the controller. 
Basic countermeasure methods such as port blocking or packet filtering may not be effective, especially in the
case of very dynamic environments (e.g. multi-tenant cloud) since connected hosts and switches change frequently which may result in preventing legitimate LLDP packets from reaching the 
controller.

\subsection{OFDP is not efficient}
By using OFDP, the controller periodically sends multiple packets to every
switch in the network, which could result in performance decrease of the data
plane. Experiments made on different controllers \cite{nectest} show that starting from
a certain network size (i.e. number of switches), running only the discovery
module results in significant increase of the controller's CPU usage and decrease
in performance.

\subsection{OFDP is not scalable}

Other issues with OFDP include that it may not 
reliably work for heavily loaded links, because discovery packets might get dropped or delayed. 
Moreover, when using OFDP in a multi-controller SDN network 
(e.g. running several tenant controllers in a virtualized network through FlowVisor), 
discovery cost increases linearly as more controllers are added.

\section{Introducing sOFTDP: Secure OpenFlow Topology Discovery Protocol}\label{introducingsOFTDP}

In a dynamic data center SDN, the controller needs to be updated whenever a topology change occurs in order to make the suitable
routing decisions for the new topology. Topology changes typically occur as a consequence of two events: 
\textit{(i)} a new link is added to the network or \textit{(ii)} an existing link is removed from the network.
Both events are the result of either adding a new switch, removing an existing switch 
or adding/removing a link between two existing switches (the latter include link and switch failures). 
sOFTDP design assumes that the controller has 
\textbf {no prior knowledge} of the occurrence of such events and it is expected to dynamically update its topology map and
adapt its routing decisions accordingly. 
%

Network as a Service (NaaS) platforms are good examples where the controller has no prior knowledge of upcoming
events. This is the case for example, of a public Cloud provider that
offers customers the possibility to create their own networks, including hosts and SDN switches in virtual machines.
The consumer (or tenant) can create hosts and switches and link them on the fly using a web interface,
while the provider's controller manages the network and ensure the connectivity. 
In this way, the tenant can add, remove, place or move 
host and switch instances without worrying about the underlying configuration.

In the remaining of this section we first identify some fundamental requirements for topology discovery 
in the context of dynamic virtualized data center networks, then we
detail the sOFTDP design choices.

\subsection{Fundamental requirements for topology discovery}
%
Topology discovery is a critical process that is required to be:

\begin{itemize}
  \item \textbf{Error free:} a topology error
  leads to wrong routing of flows. The impact can be 
  very harmful if the error is in the routing core (core routers and links)
  \item \textbf{Secure}: a discovery protocol must be secure, preventing the introduction of fake links
    and information leakage
  (including topology information).
  \item \textbf{Efficient:} a discovery protocol must not flood the controller with redundant information and only transmit the 
  topology events information when they occur.
\end{itemize}

%
%
%

\subsection{sOFTDP design}
sOFTDP \footnote{We interchangeably use the name sOFTDP for the topology discovery 
protocol and for the topology discovery application implementing it} is designed to satisfy the above requirements.
The main idea is to move a part of the discovery process from the
controller to the switch. By introducing minimal changes to the OpenFlow switch design, sOFTDP enables 
the switch to autonomously detect link events and notify the controller. We also implement
the necessary logic in the controller to handle switch notifications.
The key ingredients of sOFTDP design are: Bidirectional Forwarding Detection (BFD) as port liveness detection mechanism, 
asynchronous notifications, topology memory, FAST-FAILOVER groups, "drop lldp" rules and hashed LLDP content. 
In the following we describe each of these mechanisms. \\
\vspace{-1em}
\subsubsection{BFD as Port Liveness Detection mechanism}
  sOFTDP uses BFD (Bidirectional Forwarding Detection \cite{bfd}) as port-liveness-detection mechanism to quickly detect 
  link events. Instead of requesting topology information by sending periodic $LLDP$ frames, the controller
  just listens for link event notifications from switches to make topology updates. Hence, the switch 
  needs a mechanism to autonomously and quickly detect link events and report them to the controller.
  
  $BFD$ is a protocol that provides fast routing-protocol-independent detection of layer-3 next hop failures.
  BFD establishes a session between two 
  preconfigured endpoints over a particular link,
  and performs a control and echo message exchange to detect link
  liveliness. sOFTDP implements BFD in asynchronous mode: once 
  a session is set up with a three-way
  handshake, neighbor switches exchange periodic control messages to confirm 
  absence of a failure (presence of link) between them. Note that sOFTDP only 
  relies on BFD to detect link removal events. For link addition events, sOFTDP  
  uses OFPT\_PORT\_STATUS messages to update the topology as we will detail in the next subsection.
  The reason for using BFD instead of OFPT\_PORT\_STATUS messages 
  in detecting link removal is to include link failures that do not originate from
  administratively shutting down ports, e.g., failure of the underlying physical link or switch failure, etc.

  BFD detection time of link events depends on the control 
  packet transmission interval $T_{i}$ and the detection multiplier $M$\cite{bfd}.
  The former defines the frequency of control messages and 
  the latter defines how many control packets can be lost before the neighbor end-point
  is considered unreachable. In the worst case, 
  failure detection time is given by equation \ref{bfddetection}
  
  \begin{equation} \label{bfddetection}
      T_{det} = M * T_i 
  \end{equation}
  
  The transmit interval $T_i$ is lower-bounded by the $RTT$ of
  the link. Note that a transmit interval of $T_{i} = 16.7ms$ and 
  a detection multiplier of $M = 3$ are sufficient to achieve a detection
  time of $T_{det} = 50 ms$. Also, $M = 3$ prevents 
  small packet loss from triggering false positives. Furthermore, a such session 
  generates only $60$ packets per second. \\
  


\vspace{-1em}
  
\subsubsection{Asynchronous notifications} \label{bfdstatusmessage}
  sOFTDP enables the switch to inform the controller about port connectivity events. 
  In case of administrative changes to port status (port turned up or down), 
  the switch reports it via a OFPT\_PORT\_STATUS message defined in OpenFlow switch specifications.
  But, in the case of link failure or the remote port going down, OpenFlow 
  doesn't provide any mechanism for the switch to inform the controller. sOFTDP adds this functionality to the switch 
  by defining a new switch-to-controller message BFD\_STATUS.\\

  \vspace{-1em}
  
\subsubsection{Topology memory}
  sOFTDP keeps track of topology events and builds a database of potential backup links besides the actual link database.
  When a new link is added, sOFTDP computes the local topology (relative to the added link). If the new link forms a shorter 
  path between two switches and no traffic engineering application decides otherwise,
  the new path will be used for forwarding and the previous one will be saved as potential backup.
  sOFTDP installs OpenFlow FAST-FAILOVER groups \cite{ofspecif} 
  on the switches of the new link and marks the link 
  as 'safe to remove' since it has at least one potential backup. Note that potential backup links are not considered 
  backup links until all traffic engineering applications agree.
  Traffic engineering applications must communicate with sOFTDP to prevent 
  interference in selecting primary paths and backups.\\
  
  \vspace{-1em}
  
\subsubsection{FAST-FAILOVER groups}
  OpenFlow groups enable OpenFlow to abstract a set of ports as a single forwarding entity 
  allowing advanced forwarding and monitoring at the switch level. 
  The group table contains group entries; each group entry is composed of a set of action buckets with specific
  semantics dependent on the group type. When a packet is sent to a group, 
  the actions in one or more action buckets are applied to it before forwarding to the egress port.
  Groups buckets can also forward to other groups, enabling to chain groups together.
  
  The following four types of group tables are provided: 
  
  \begin{itemize}
   \item \textbf{All:} used for multicast and  flooding
   \item \textbf{Select:} used for multipath
   \item \textbf{Indirect:} simple indirection
   \item \textbf{Fast Failover:} use  first live port
  \end{itemize}

  Different types of group tables are associated with different abstractions such as multicasting
  or multipathing. In particular, the Fast Failover Group
  Table monitors the status of ports and applies forwarding actions accordingly.
  When the monitored port goes down, the Fast Failover Group Table switches
  to the first port alive without consulting the controller \cite{ofspecif}.
  
  sOFTDP enables seamless removal of switches and links while preserving connectivity. 
  In order to accomplish that, when a link removal event occurs, 
  sOFTDP uses OpenFlow FAST-FAILOVER groups (optional in OpenFlow 1.1+) to watch switch ports and 
  perform fast switchover to backup links. Hence, switches concerned by the link removal 
  start forwarding flows through the backup link and
  do not have to wait until the controller receives the topology event and installs new rules.  \\

%

  \vspace{-1em}
  
\subsubsection{"drop lldp" rules}
  The switch has a rule "drop lldp" to drop every LLDP packet to
  prevent LLDP flooding attacks (see figure \ref{attacks.c}). In a SDN running OFDP, 
  traditional Denial of Service (DoS) mitigating methods like placing firewalls or Intrusion Detection Systems (IDSs) 
  to filter out LLDP packets are not effective because it is hard to distinguish between legitimate LLDP packets 
  (generated by the controller and forwarded by switches) from the fake ones (generated by the attacker and also forwarded by 
  switches to the controller). By removing periodically broadcasted LLDP packets, sOFTDP eliminates 
  the possibility that malicious LLDP packets get forwarded to the controller and hence prevents it from being flooded.\\
  
   \vspace{-1em}
   
\subsubsection{Hashed LLDP content}
  The controller sends encrypted LLDP packets only when it receives a OFPT\_PORT\_STATUS with 
  the flag $PORT\_UP$ set to $1$ indicating the port went from down to up status. 
  The LLDP packets are sent only to the concerned
  switches along with OpenFlow rules to forward them to the controller.
  These rules must have a higher priority
  than "drop lldp" rules and their $hard timeout$ values are set to $500ms$.
  The purpose of the LLDP packets here is to learn added links as shown in figure 
  \ref{howworks.b} and detailed in subsection \ref{howsoftdpworks}.
  Finally, $500ms$ is a small arbitrary value to ensure 
  that potential malicious LLDP packets generated exactly during this time window will not significantly affect the controller.\\
  
  \vspace{-1em}

\subsection{How sOFTDP works} \label{howsoftdpworks}

\begin{figure}[h] 
    \centering
  \label{fig7} 
  \begin{subfigure}[b]{0.5\linewidth}
    \centering
    \includegraphics[width=1\linewidth]{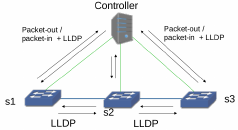}
    \caption{\label{howworks.a}Bootstrap} 
  \end{subfigure}
  \hspace{1ex}
  \begin{subfigure}[b]{0.45\linewidth}
    \centering
    \includegraphics[width=1\linewidth]{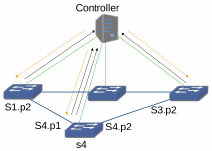}
    \caption{\label{howworks.b}s4 joins the network} 
  \end{subfigure} 
  \begin{subfigure}[b]{0.47\linewidth}
    \centering
    \includegraphics[width=1\linewidth]{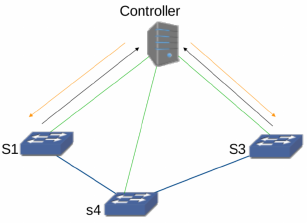}
    \caption{\label{howworks.c}s2 leaves the network} 
  \end{subfigure}
 \hspace{1ex}
  \begin{subfigure}[b]{0.45\linewidth}
    \centering
    \includegraphics[width=1\linewidth]{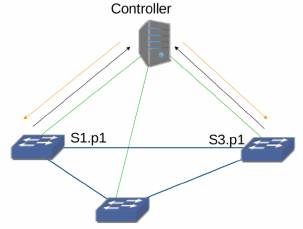}
    \caption{\label{howworks.d}Addition of link $s1 \leftrightarrow s3$} 
  \end{subfigure} 
  \begin{subfigure}[b]{0.45\linewidth}
    \centering
    \includegraphics[width=1\linewidth]{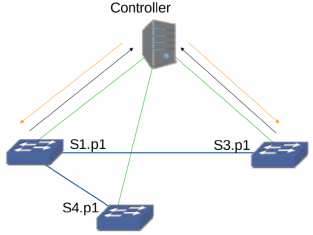}
    \caption{\label{howworks.e}Removal of link $s3 \leftrightarrow s4$} 
  \end{subfigure}  
  \caption{\label{howworks}How sOFTDP works}
\end{figure}

Figure \ref{howworks} shows how sOFTDP works. To bootstrap, the controller sends LLDP packets to all connected switches  
like in traditional OFDP (figure \ref{howworks.a}).
The main difference is that we do not use clear MAC addresses as switch DPIDs. Instead, 
we use hash values of them to prevent all 
information discloser and switch spoofing attacks. We also hash \textit{system\_description} field value to prevent
controller fingerprinting \cite{sdnfingerprint}.

When a new switch joins the network, it starts by establishing a connection with the controller:
The switch and the controller exchange $hello$ $(OFPT\_HELLO)$ 
messages that specify the latest OpenFlow protocol version supported by the sender.
Then, the controller sends the switch a $feature\ request$ message asking its capabilities. 
The switch responds with a $feature\ reply$ message, which includes the local MAC address 
(that corresponds to the internal port of the switch) in the 
$switch\ datapath\ ID$ field and that's how the controller keeps track of connected switches. 
The $feature\ reply$ message also includes
ports status which are all down initially. In figure \ref{howworks.b}, 
when switch $s4$ joins the network and new links are established,
switches s1, s3 and s4 send 
$PORT\_STATUS$ messages to inform the controller that involved ports s1.p2, s4.p1, s4.p2 and s3.p2 went up and are connected.
The controller 'c' then sends LLDP packets to be forwarded only through those ports: 
$c \rightarrow s1.i \rightarrow s1.p2 \rightarrow s4.p1 \rightarrow s4.i \rightarrow c$,
$c \rightarrow s4.i \rightarrow s4.p2 \rightarrow s3.p2 \rightarrow s3.i \rightarrow c$, 
$c \rightarrow s3.i \rightarrow s3.p2 \rightarrow s4.p2 \rightarrow s4.i \rightarrow c$ and 
$c \rightarrow s4.i \rightarrow s4.p1 \rightarrow s1.p2 \rightarrow s1.i \rightarrow c$. With i indicates internal port.

Once all LLDP packets are received, the controller identifies the new links and store them in the Topology Map.
Note that by using $PORT\_STATUS$ messages as trigger to learn new links, the controller doesn't need to periodically 
send discovery packets and switches do not need to be too smart to determine the new links (as in \cite{optimalforce})
or to store them locally.

Once the new topology is computed, the controller detects multiple paths between pairs of switches.
Independently of traffic engineering applications running on the same controller, the sOFTDP topology module tags shortest paths 
as primary paths and longer paths as secondary paths or potential backups 
(e.g., $s1 \leftrightarrow s2$ is a primary path and $s1 \leftrightarrow s4 \leftrightarrow s3 \leftrightarrow s2$ is a secondary path). 
Then, if not specified otherwise by any traffic engineering application, 
the controller installs fast-failover group rules on the switches of the shortest path. 
This ensures continuity of connectivity in case of topology events. In the example shown in figure \ref{howworks.b}, 
there are two similar paths, in term of number of hops, between $s1$ and $s3$.
The controller arbitrary tags $s1 \leftrightarrow s4 \leftrightarrow s3$ as primary path and 
installs fast-failover group rules on switches $s1$ and $s3$ to watch ports $s1.p1$, $s1.p2$, $s3.p1$ and $s3.p2$.


When a switch leaves the network ($s2$ in figure \ref{howworks.c}), neighbor switches detect and report link events to the controller:
BFD session on $s1.p1$ detects the link $s1.p1 \leftrightarrow s2.p1$ failure and sends a BFD\_STATUS message to the controller. 
In the case of link removal, the controller doesn't need to send LLDP packets and just removes the link from the topology map.
The same process applies to link $s2.p2 \leftrightarrow s3.p1$.
Switches s1 and s3 automatically switch traffic through the path s1 $\leftrightarrow$ s4 $\leftrightarrow$ s3 using the 
fast-failover group rules installed previously.

When a link is added between two existing switches ($s1 \leftrightarrow s3$ in figure \ref{howworks.d}), 
the involved ports $s1.p1$ and $s3.p1$
send $PORT\_STATUS$ messages to the controller with "port up" flags set. 
The controller then sends LLDP packets to be forwarded only through
$s1.p1$ and $s3.p1$: $controller \rightarrow s1.internal \rightarrow s1.p1 \rightarrow s3.p1 \rightarrow controller$ and 
$controller \rightarrow s3.internal \rightarrow s3.p1 \rightarrow s1.p1 \rightarrow controller$.
After the new topology is computed, the controller tags $s1 \leftrightarrow s3$ as the shortest path 
and $s1 \leftrightarrow s4 \leftrightarrow s3$ as a potential backup path and installs 
fast-failover group rules on $s1$ and $s3$ (in this particular example, the same rules already exist)

When an existing link is removed ($s3 \leftrightarrow s4$ in figure \ref{howworks.e}), the involved ports $s3.p2$ 
and $s4.p2$ detect loss of connectivity very quickly using BFD and report it to the controller via BFD\_STATUS messages. 
The controller then drops the link $s3 \leftrightarrow s4$ from its topology map without the need to send LLDP packets. 
Finally, the controller removes the tag from the remaining path.

\section{Evaluation} \label{Evaluation}
\subsection{Emulation Testbed}

To evaluate sOFTDP, we implemented sOFTDP topology module on Floodlight.
We conducted 
experiments on an emulated testbed using Mininet \cite{mininet}. 
The emulated testbed is composed of four virtual bridges based on Open vSwitch \cite{vswitch} 
and controlled by 
Floodlight controller from a different physical machine.
We upgraded existing Open vSwitch (of mininet v2.2.1) to the newer 
version 2.3.1 that supports the BFD protocol and fast failover groups.
Then we added a simple patch to Open vSwitch to send BFD\_STATUS to 
the controller upon BFD events (see section \ref{bfdstatusmessage}).
BFD detection time is set to $1ms$. 

%
%



\subsection{Experiments and results}

\begin{figure*}[ht] 
\centering
  \label{ fig7} 
  \begin{subfigure}[b]{0.3\linewidth}
    \centering
    \includegraphics[width=1\linewidth]{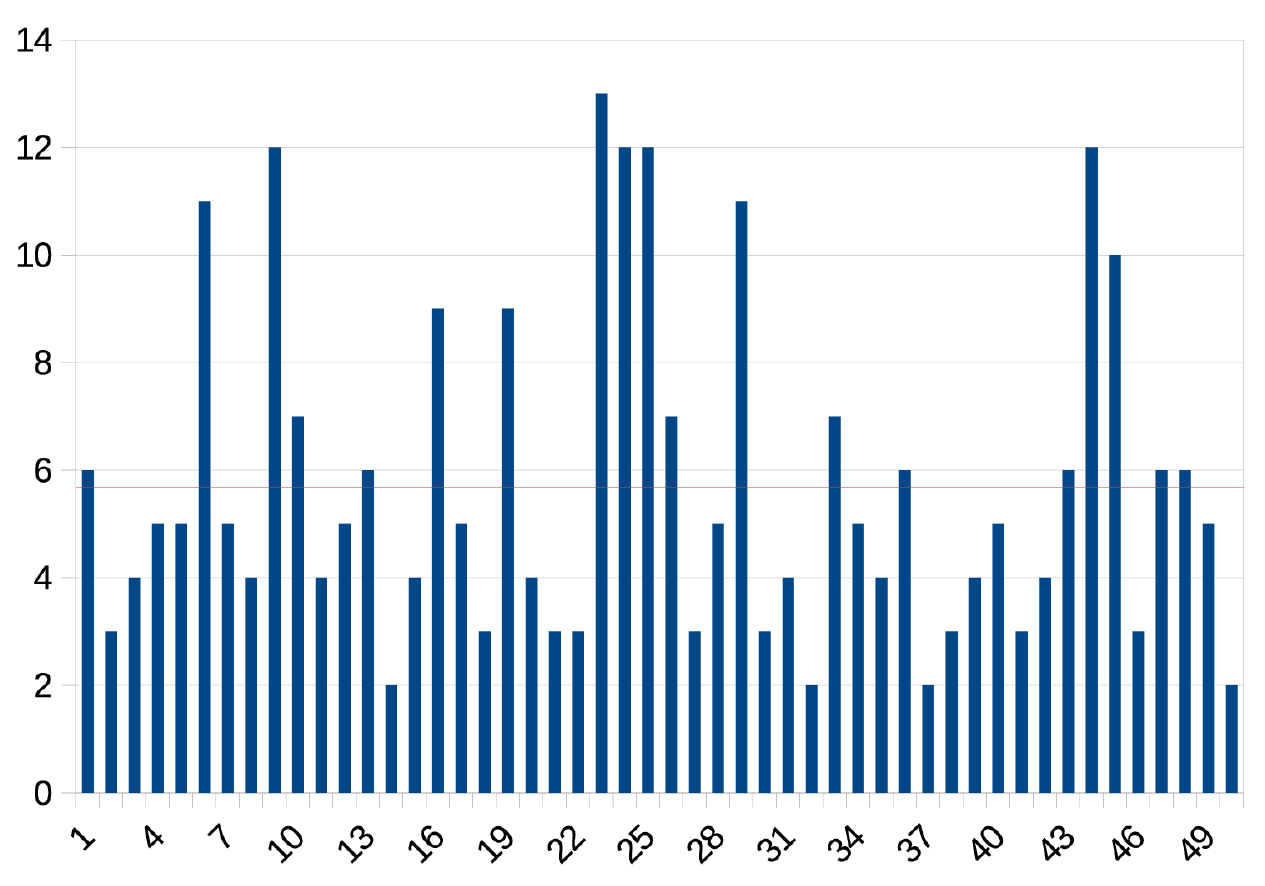}
    \caption{\label{results.a}New link detection time in ms} 
  \end{subfigure}
  \begin{subfigure}[b]{0.3\linewidth}
    \centering
    \includegraphics[width=1\linewidth]{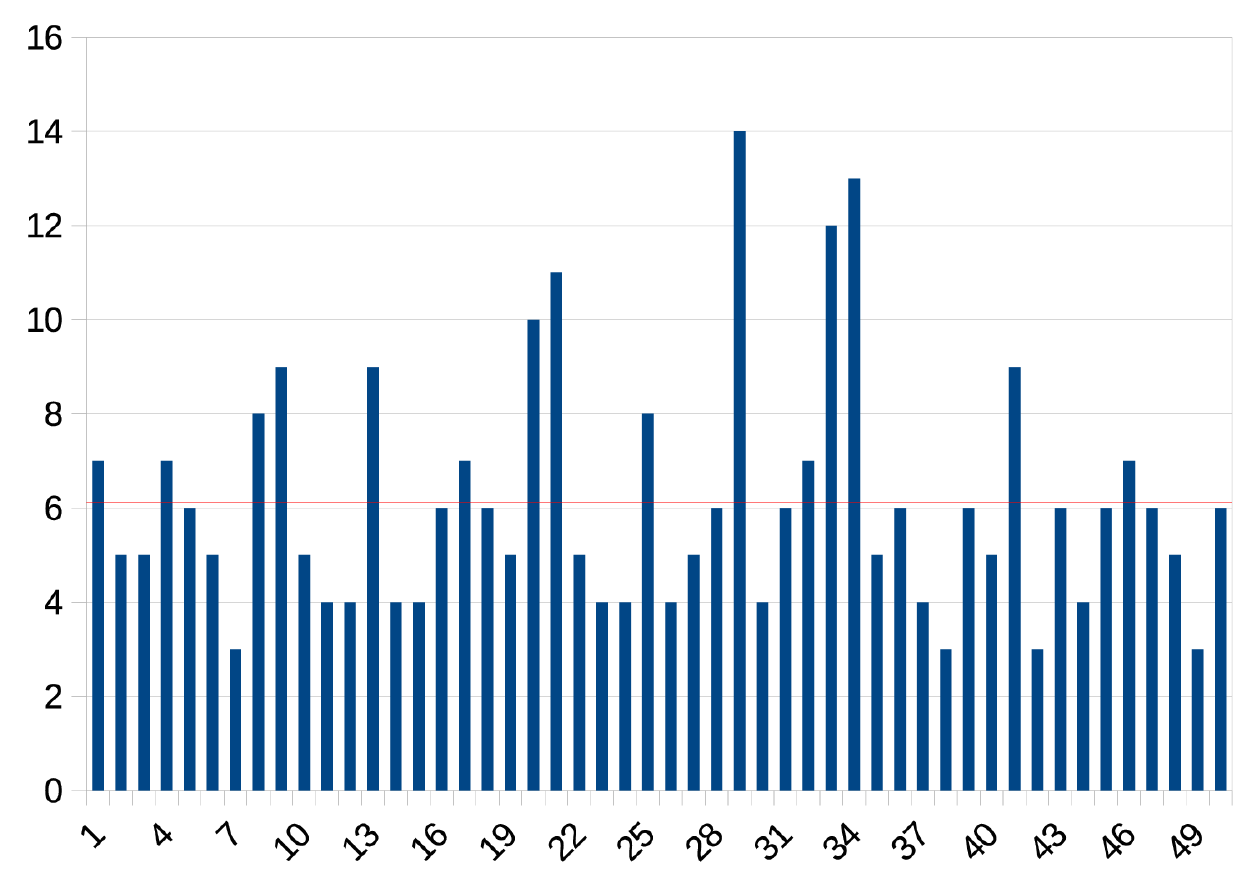}
    \caption{\label{results.b}Adaptation time in ms} 
  \end{subfigure} 
  \begin{subfigure}[b]{0.3\linewidth}
    \centering
    \includegraphics[width=1\linewidth]{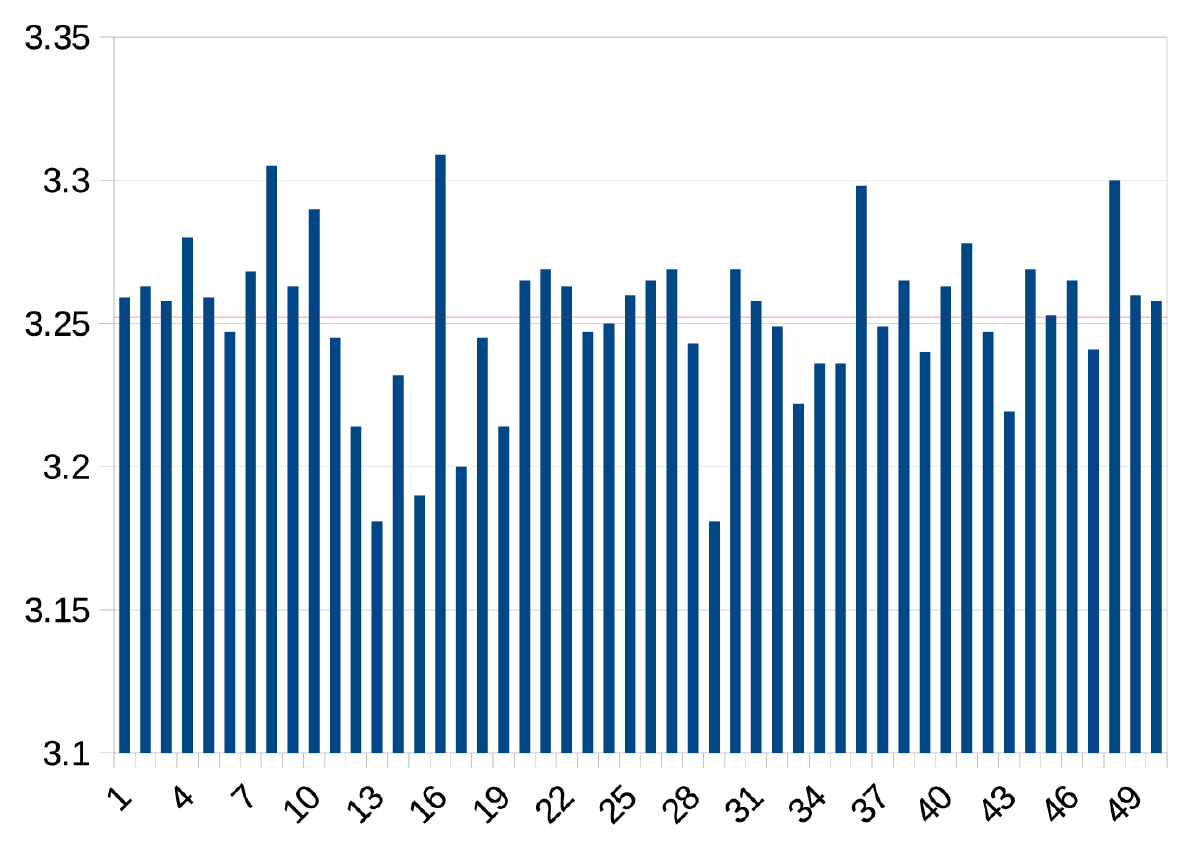}
    \caption{\label{results.c}Link removal detection time in ms} 
  \end{subfigure}
  
  \caption{\label{results}Evaluation of sOFTDP}
  \vspace{-1em}
\end{figure*}

%
%

As previously explained, to handle link removal, sOFTDP uses BFD protocol to detect port events at the switch level, 
then the switch triggers a notification the the controller 
(BFD\_STATUS message) and finally the controller removes the link from the topology map.
To handle link addition, sOFTDP listens for OFPT\_PORT\_STATUS messages triggered by the ports 
going up then the controller sends a LLDP message to the concerned 
switches to confirm the added link and finally adds it to the topology map.
Accordingly, two scenarios are implemented and evaluated:

\textbf{Scenario one:} 
Link $s1.p1 \leftrightarrow s3.p1$ is added (figure \ref{howworks.d}). 
We measure the $learning\ time$  the controller takes to know about the added link as given in equation \ref{learningtime}.

\begin{equation} \label{learningtime}
    T_{learn}(i,j) = max_{d\in\{i,j\}}(T_{pstatus}(d)) + RTT_{LLDP}(i , j)
\end{equation}
\begin{equation}
    RTT_{LLDP}(i,j) = T_{delv}(c, i) + T_{delv}(i, j) + T_{delv}(j, c)
\end{equation}
\begin{equation} 
    T_{learn}(\{i,j\}) = max(T_{learn}(i,j), T_{learn}(j,i))
\end{equation}
Where: $T_{learn}(i,j)$ is the time necessary to learn unidirectional link $(i,j)$
$T_{pstatus}(i) = T_{trsm}(i, c)$ is the time OFPT\_PORT\_STATUS message takes from switch $i$ to the controller. 
$RTT_{LLDP}(i,j)$ is the round trip time that a LLDP packet sent from the controller takes to go through switch i then switch j and
back to the controller.
$T_{learn}(\{i,j\})$ is the time necessary to learn bidirectional link $\{(i,j),(j,i)\}$ and finally $T_{delv}(x, y)$ is
the packet delivery time from node $x$ to node $y$.
Figure \ref{results.a} shows the average of 50 performed experiments and $95\%$ confidence interval yielding learning 
times of $5.68 \pm 0.85ms $

To further demonstrate sOFTDP performance, we also measure the overall $adaptation\ time$ 
when a new link is added (see figure \ref{howworks.d}) to the network.
$adaptation\ time$ includes learning the new link, installing fast-failover group rules in the switches and 
the actual switchover time.
Figure \ref{results.b} depicts the result averaged from 50 conducted experiments. The average and $95\%$ confidence interval 
are of $6.12\pm0.7ms$.

  \begin{figure} 
\centering
   \includegraphics[scale=0.35]{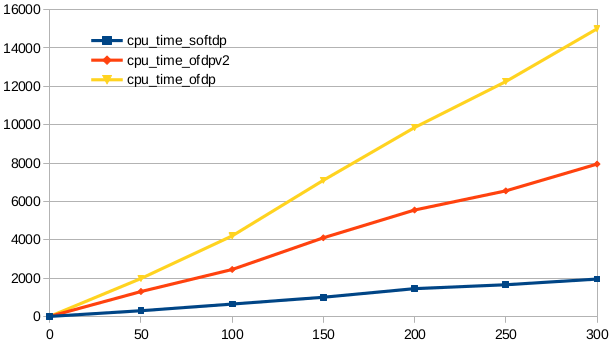}
   \caption{\label{fig:comparison} CPU time (in ms) over number of switches}
   \vspace{-1em}
\end{figure} 

\textbf{Scenario two:}
Link $s3.p2 \leftrightarrow s4.p2$ is removed (figure \ref{howworks.e}).
We measure the $learning\ time$  the controller takes to learn the change in the topology. 
The link is brought down
upon the first BFD\_STATUS message of either of its end­points.

The $learning\ time$ in this case is:

\begin{equation} \label{learningtime2}
    T_{learn}(i,j) = min(T_{bfd}(i), T_{bfd}(j)) 
\end{equation}

Where $T_{bfd}(x)$ is the BFD detection time on the involved port of node $x$. Computed as follows:

\begin{equation} \label{learningtime3}
    T_{bfd}(x) = T_{det}^x + T_{delv}(x,c) 
\end{equation}
Where $T_{det}^x$ is $BFD\ detection\ time$ given in equation \ref{bfddetection} and $T_{delv}(x,c)$ is $packet\ delivery\ time$
from node $x$ to the controller.

Figure \ref{results.c} shows sOFTDP learning time average (taken over 50 performed experiments) and the $95\%$
confidence interval resulting in $3.25\pm0.008ms$

sOFTDP $learning\ time$ is independent of the size of the network and depends only on the inter-switch
$RTT$ and the $RTT$ between switches and the controller. Figure \ref{fig:comparison} 
compares sOFTDP to OFDP and OFDPv2 \cite{simplepaper} in term of CPU time. Each experiment is performed over $200s$ period during which 
one topology event is generated every second.

%

%
%
%
%
%
%
%
%
%

\section{Related work} \label{relatedwork}

Unlike our proposal, most of previous work focus either on security problems or on performance problems in OFDP.
In \cite{poisonning}, authors identified $link\ fabrication$ attacks on OFDP and proposed to authenticate LLDP packets by adding
an optional TLV "HMAC" to ensure their origin and integrity.

The average overhead introduced by this approach differs between the first discovery round and the following rounds, because  the 
HMAC value is computed once and cached for the future construction and validation of LLDP packets.
The average overhead accounts for $80.4\%$ of overall LLDP construction time in the first round and accounts for $2.92\%$
in the following rounds. Although this approach prevents link fabrication by 
fake LLDP injection, it does not defend against link fabrication in a relay manner as discussed in section \ref{whyofdpbad}.
The authors argue that a solution to the link fabrication by a relay could be that the controller monitors ports to detect 
whether the connected machine is a host or a switch. If the connected machine is a host generating LLDP packets then an alert is 
triggered. However, a host can easily behave like a switch making this solution unpractical. 

A similar approach was proposed in \cite{insecurity} but using HMAC with a dynamic key which is randomly generated for every
single LLDP packet. This approach adds an extra $8\%$ of in CPU load.

OFDPv2 \cite{simplepaper} reduces the number of OFDP-related packet-out messages by rewriting LLDP packet headers in the switch. 
In the traditional OFDP, the controller sends $\sum_{i=1}^{n} p_i$ packet-out messages every discovery round, 
where $n$ is the number of switches and $p_i$ the number of ports in switch $i$. The number of packet-out messages shrinks to $n$
by sending only one packet per switch and rewriting copies for different ports at level of the switch. OFDPv2 achieves $50\%$
reduction in CPU load compared to OFDP but obviously requires more logic to be added to the switch. 
Also, OFDPv2 does not reduce the number of packet-in messages that the controller periodically receives from switches.

In \cite{optimalforce} authors implemented the ForCES \cite{forces} protocol to communicate the
topology information between switches and the controller. Switches acquire neighbor topology information by exchanging LLDP
packets as in traditional networks and store it in their device maps. The acquired information is updated 
periodically as LLDP frames are exchanged. Then, upon receiving a topology change notification from a switch,
the controller needs to query the connected switches in order to learn their respective neighbors. The authors measured 
an average learning time of $12ms$ without considering the LLDP exchange time. In other words, the LLDP time exchange time is not 
included and it takes $12ms$ for the switch to detect the topology change, send a notification to the controller
and then answer the controller's request for the topology information.
%

\section{Conclusion}\label{conclusion}

In this work, we extended our previous paper on OFDP limitations by introducing and detailing a novel topology discovery protocol
for OpenFlow (sOFTDP). We argue that this is the first time major security and performance issues related to the topology discovery process 
in current SDN controllers, are tackled. Our proposal requires minimal changes to the OpenFlow switch design and is shown to be 
more secure (by design) than previous workarounds on traditional OFDP. Also, our proposal outperforms OFDP and OFDPv2 by several orders of 
magnitude which we confirmed by proof of concept experiments. Further experiments on larger physical testbeds are being conducted ad will
be included in future work.

%
%

\end{document}